\documentclass{sf2a-conf2010}
\usepackage{amssymb}
\usepackage{graphicx}
\usepackage{hyperref}
\usepackage[]{natbib}

\begin{document}
\TitreGlobal{SF2A 2010}
%
\title{Constraints on the cosmic ray diffusion coefficient \\ in the W28 region from gamma-ray observations}
\author{S. Gabici}\address{Laboratoire APC, 10 rue A Domon et L Duquet, 75205 Paris Cedex 13, France}
\author{S. Casanova$^{2,}$}\address{Max-Planck-Institut f\"ur Kernphysik, Heidelberg, Germany}\address{Ruhr-Universitaet Bochum, Fakultaet fuer Physik und Astronomie, Bochum, Germany}
\author{F. A. Aharonian$^{2,}$}\address{Dublin Institute for Advanced Studies, Dublin, Ireland}
\author{G. Rowell}\address{School of Chemistry and Physics, University of Adelaide, Adelaide, Australia}
\runningtitle{Cosmic ray diffusion in the W28 region}
%
\setcounter{page}{237}
\index{Gabici, S.}
\index{Casanova, S.}
\index{Aharonian, F. A.}
\index{Rowell, G.}

\maketitle
\begin{abstract}
%
%
GeV and TeV gamma rays have been detected from the supernova remnant W28 and its surroundings. Such emission correlates quite well with the position of dense and massive molecular clouds and thus it is often interpreted as the result of hadronic cosmic ray interactions in the dense gas. Constraints on the cosmic ray diffusion coefficient in the region can be obtained, under the assumption that the cosmic rays responsible for the gamma ray emission have been accelerated in the past at the supernova remnant shock, and subsequently escaped in the surrounding medium. In this scenario, gamma ray observations can be explained only if the diffusion coefficient in the region surrounding the supernova remnant is significantly suppressed with respect to the average galactic one.  
\end{abstract}
\begin{keywords}
cosmic rays, gamma rays, supernova remnants
\end{keywords}
%

Cosmic ray (CR) spallation measurements allow us to infer the average residence time $t_{res}$ of CRs in the Galaxy. Then, if $h$ is the distance that a CR has to move away from its source before escaping the Galaxy (i.e. the Galaxy's thickness), it is possible to estimate the CR diffusion coefficient as $D \approx h^2/t_{res}$. A thorough comparison between spallation data and CR transport models gives: $D_{gal} \approx 10^{28} (E/10~{\rm GeV})^{\delta}~{\rm cm^2 s^{-1}}$, where $E$ is the CR particle energy and $\delta \sim 0.3 \div 0.7$ \citep{CRbook}. However, this has to be intended as the average CR diffusion coefficient in the Galaxy, and local variations (both in time and space) of the diffusion coefficient might exist.

In particular, the diffusion coefficient might be suppressed close to CR sources. This is because CRs can excite magnetic turbulence while streaming away from their acceleration site. This would enhance the scattering rate of CR themselves and thus reduce the diffusion coefficient \citep{kulsrud}. The problem of estimating, on theoretical grounds, the diffusion coefficient around CR sources is far from being solved, mainly because of its intrinsic non-linearity and because various mechanisms might damp the CR--generated waves and thus affect the way in which CRs diffuse  \citep[e.g.][]{farmer,ptuskin}. 

Gamma ray observations can provide us with constraints on the diffusion coefficient close to CR sources. After escaping from their sources, CRs undergo hadronic interactions with the surrounding gas and produce gamma rays. The characteristics of such radiation (i.e. its spectrum and intensity as a function of the time elapsed since CRs escaped the source) depend on the value of the diffusion coefficient that can thus be constrained, if a reliable modeling for CR acceleration at the source is available. The presence of massive Molecular Clouds (MCs) close to the source would enhance the gamma ray emission, making its detection more probable.

The study of gamma ray emission from runaway CRs has been studied by \citet{atoyan} for the case of a generic CR source, while the specific (and most popular) situation of CR acceleration at SuperNova Remnants (SNRs) has been described in a series of recent papers \citep{gabici2007,gabici2009,casanova2010}. The study of such radiation is of great importance not only in order to reach a better understanding of the way in which CRs diffuse in the interstellar magnetic field, but also because its detection can provide an indirect way to identify the sources of galactic CRs.

W28 is an old SNR in its radiative phase of evolution, located in a region rich of dense molecular gas with average density $\gtrsim 5~{\rm cm^{-3}}$. At an estimated distance of $\sim 2~{\rm kpc}$ the SNR shock radius is $\sim 12~{\rm pc}$ and its velocity $\sim 80~{\rm km/s}$ \citep[e.g.][]{rho}. By using the dynamical model by \citet{cioffi} and assuming that the mass of the supernova ejecta is $\sim 1.4 ~ M_{\odot}$, it is possible to infer the supernova explosion energy ($E_{SN} \sim 0.4 \times 10^{51} {\rm erg}$), initial velocity ($\sim 5500~{\rm km/s}$), and age ($t_{age} \sim 4.4 \times 10^4 {\rm yr}$).

Gamma ray emission has been detected from the region surrounding W28 both at TeV \citep{hess} and GeV energies \citep{fermi,agile}, by HESS, FERMI, and AGILE, respectively. The TeV emission correlates quite well with the position of three massive molecular clouds, one of which is interacting with the north-eastern part of the shell (and corresponds to the TeV source HESS J1801-233), and the other two being located to the south of the SNR (TeV sources HESS J1800-240 A and B) . The masses of these clouds can be estimated from CO measurements and result in $\approx 5$, $6$, and $4 \times 10^4 M_{\odot}$, respectively, and their projected distances from the centre of the SNR are $\approx$ 12, 20, and 20 pc, respectively \citep{hess}. The GeV emission roughly mimics the TeV one, except for the fact that no significant emission is detected at the position of HESS J1800-240 A. 

In this paper, we investigate the possibility that the gamma ray emission from the W28 region could be the result of hadronic interactions of CRs that have been accelerated in the past at the SNR shock and then escaped in the surrounding medium \footnote{This scenario has been described in a number of recent papers \citep{fujita,ohira,li}.}. To do so, we follow the approach described in \citet{gabici2009} (hereafter GAC2009), which we briefly summarize below. 

For each particle energy $E$ we solve the diffusion equation for CRs escaping the SNR. For simplicity we treat the SNR as a point like source of CRs and we consider an isotropic and homogeneous diffusion coefficient $D = \chi D_{gal} \propto E^{0.5}$ (see Eq. 5 in GAC2009). Here, $\chi$ represents possible deviations with respect to the average CR diffusion coefficient in the Galaxy. The solution of the diffusion equation gives the spatial distribution of CRs around the source $f_{CR}$, which is roughly constant up to a distance equal to the diffusion radius $R_d = \sqrt{4 ~ D ~ t_{diff}}$, and given by $f_{CR} \propto \eta E_{SN}/R_d^3$, where $\eta$ is the fraction of the supernova explosion energy converted into CRs, and $t_{diff}$ is the time elapsed since CRs with energy $E$ escaped the SNR (see eq. 6 in GAC2009). For distances much larger than $R_d$ the CR spatial distribution falls like $f_{CR} \propto \exp(-(R/R_d)^2)$ (see eq. 6 in GAC2009). Following GAC2009, we assume that CRs with energy 5 PeV (1 GeV) escape the SNR at the beginning (end) of the Sedov phase, at a time $\sim 250 ~ {\rm yr}$ ($\sim 1.2 \times 10^4 ~{\rm yr}$) after the explosion, and that the time integrated CR spectrum injected in the interstellar medium is $\propto E^{-2}$. In this scenario, particles with lower and lower energies are released gradually in the interstellar medium \citep{vladvlad}, and we parametrize the escape time as: $t_{esc} \propto E^{-\alpha}$ which, during the Sedov phase, can also be written as $R_s \propto E^{-2 \alpha /5}$, where $R_s$ is the shock radius at time $t_{esc}$ and $\alpha \sim 4$. From this it follows that the assumption of point like CR source is good for high energy CRs only ($\sim$ TeV or above), when $R_s$ is small, but it becomes a rough approximation at significantly lower energies. This is because low energy particles are believed to be released later in time, when the SNR shock radius is large (i.e. non negligible when compared to $R_d$). 

We now provide a simplified argument to show how we can attempt to constrain the diffusion coefficient by using the TeV gamma ray observations of the MCs in the W28 region. The time elapsed since CRs with a given energy escaped the SNR can be written as: $t_{diff} = t_{age} - t_{esc}$. However, for CRs with energies above 1~TeV (the ones responsible for the emission detected by HESS) we may assume $t_{esc} << t_{age}$ (i.e. high energy CRs are released when the SNR is much younger than it is now) and thus $t_{diff} \sim t_{age}$. Thus, the diffusion radius reduces to $R_d \sim \sqrt{4 ~ D ~ t_{age}}$. We recall that within the diffusion radius the spatial distribution of CRs, $f_{CR}$, is roughly constant, and proportional to $\eta E_{SN}/R_d^3$. On the other hand, the observed gamma ray flux from each one of the MCs is: $F_{\gamma} \propto f_{CR} M_{cl}/d^2$, where $M_{cl}$ is the mass of the MC and $d$ is the distance of the system. Note that in this expression $F_{\gamma}$ is calculated at a photon energy $E_{\gamma}$, while $f_{CR}$ is calculated at a CR energy $E_{CR} \sim 10 \times E_{\gamma}$, to account for the inelasticity of proton-proton interactions. By using the definitions of $f_{CR}$ and $R_d$ we can finally write the approximate equation, valid within a distance $R_d$ from the SNR:
$$
F_{\gamma} \propto \frac{\eta ~ E_{SN}}{(\chi ~ D_{gal} ~ t_{age})^{3/2}} \left( \frac{M_{cl}}{d^2} \right) .
$$

Estimates can be obtained for all the physical quantities in the equation except for the CR acceleration efficiency $\eta$ and the local diffusion coefficient $\chi D_{gal}$. By fitting the TeV data we can thus attempt to constrain, within the uncertainties given by the errors on the other measured quantities (namely, $E_{SN}$, $t_{age}$, $M_{cl}$, and $d$) and by the assumptions made (e.g. the CR injection spectrum is assumed to be $E^{-2}$), a combination of these two parameters (namely $\eta/\chi^{3/2}$). The fact that the MCs have to be located within a distance $R_d$ from the SNR can be verified a posteriori. Given all the uncertainties above, our results have to be interpreted as a proof of concept of the fact that gamma ray observations of SNR/MC associations can serve as tools to estimate the CR diffusion coefficient. More detection of SNR/MC associations are needed in order to check whether the scenario described in this paper applies to a whole class of objects and not only to a test-case as W28. Future observations from the Cherenkov Telescope Array will most likely solve this issue.

\begin{figure}[t!]
 \centering
 \includegraphics[width=0.5\textwidth]{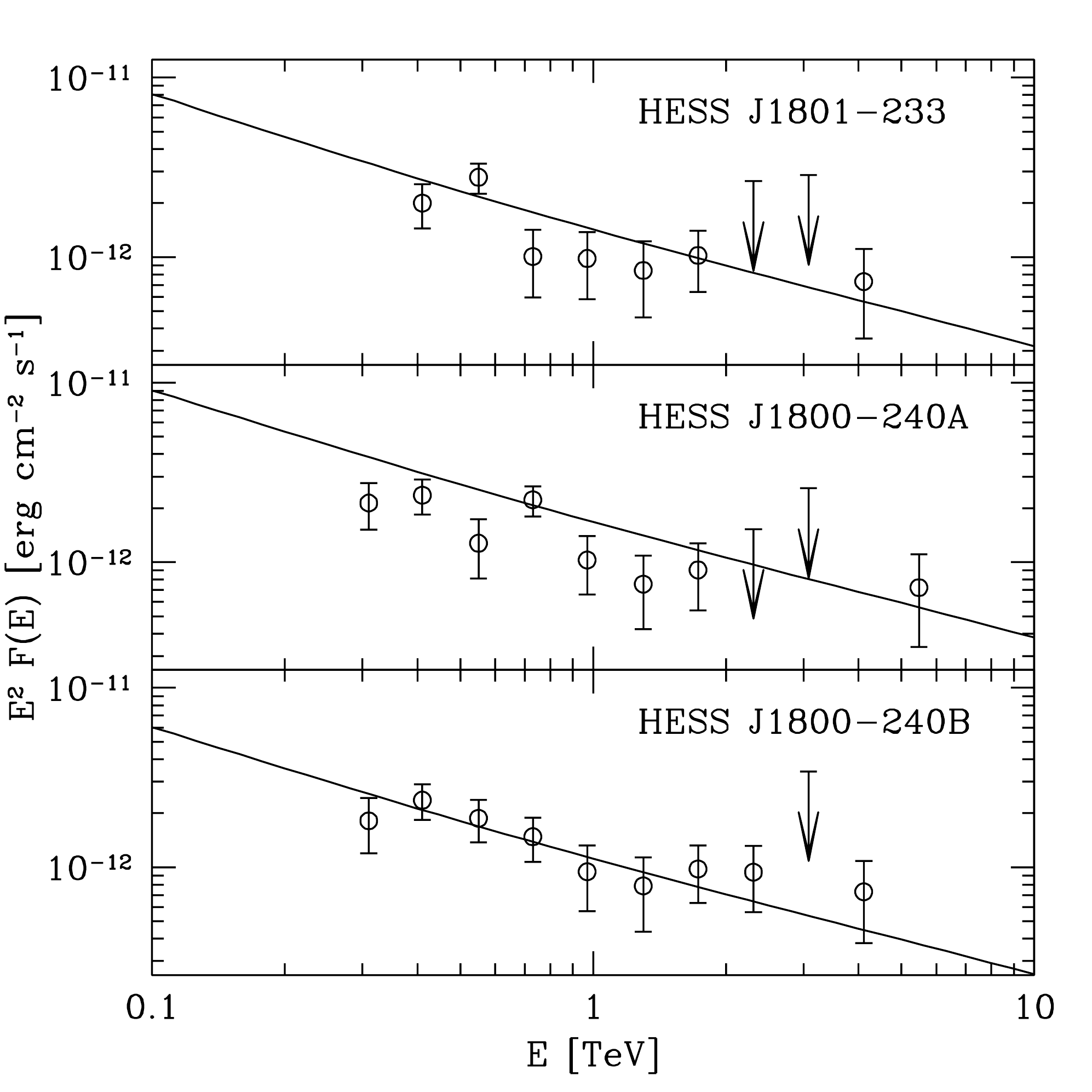}      
  \caption{Simultaneous fit to the three TeV sources detected by HESS in the W28 region. Gamma ray spectra have been calculated by using the parameterizations by \citet{kamae}, where a multiplicative factor of 1.5 has been applied to account for the contribution to the emission from nuclei heavier than H both in CRs and in the interstellar medium. }
  \label{fig1}
\end{figure}

Fig. \ref{fig1} shows a fit to the HESS data for the three massive MCs in the W28 region. A simultaneous fit to all the three MCs is obtained by setting  $\eta/\chi^{3/2} \approx 20$, which implies that the diffusion coefficient (normalized to the average galactic one) $\chi$ has to be much smaller than 1 for any reasonable value of $\eta < 1$. For example, an acceleration efficiency $\eta = 30\%$ corresponds to a CR diffusion coefficient of $\chi = 0.06$, which in turn gives a diffusion distance for TeV particles of $R_d \approx 60~{\rm pc}$. This means that the results in Fig.~\ref{fig1} are valid if the physical (not projected) distances between the MCs and the SNRs do not significantly exceed $R_d$.  Small values of the diffusion coefficient have been also proposed by \citet{agile,fujita,li}. Note that, since we are considering gamma rays in a quite narrow (about one order of magnitude) energy band around $\approx 1$~TeV, we are actually constraining the diffusion coefficient of CRs with energy $\approx 10$~TeV, and we cannot say much about the energy dependence of the diffusion coefficient 
on a broad energy interval. 

In principle, observations by FERMI and AGILE might be used to constrain the diffusion coefficient down to GeV particle energies. However, in this energy band the uncertainties are more severe because of the following reasons: {\it i)} low energy CRs are believed to be released late in time, when the SNR shock is large, and thus the assumption of point-like source is probably not well justified (see \citealt{ohira} for a model that takes into account the finite size of the SNR) ; {\it ii)} for the same reason, we can no longer assume that $t_{diff} \sim t_{age}$, as we did for high energy CRs. In other words, we need to know the exact time at which CRs with a given energy escape the SNR. Though some promising theoretical studies exist \citep[e.g.][]{vladvlad}, our knowledge of the escape time of CRs from SNRs is still quite limited.

\begin{figure}[t!]
 \centering
 \includegraphics[width=0.3\textwidth]{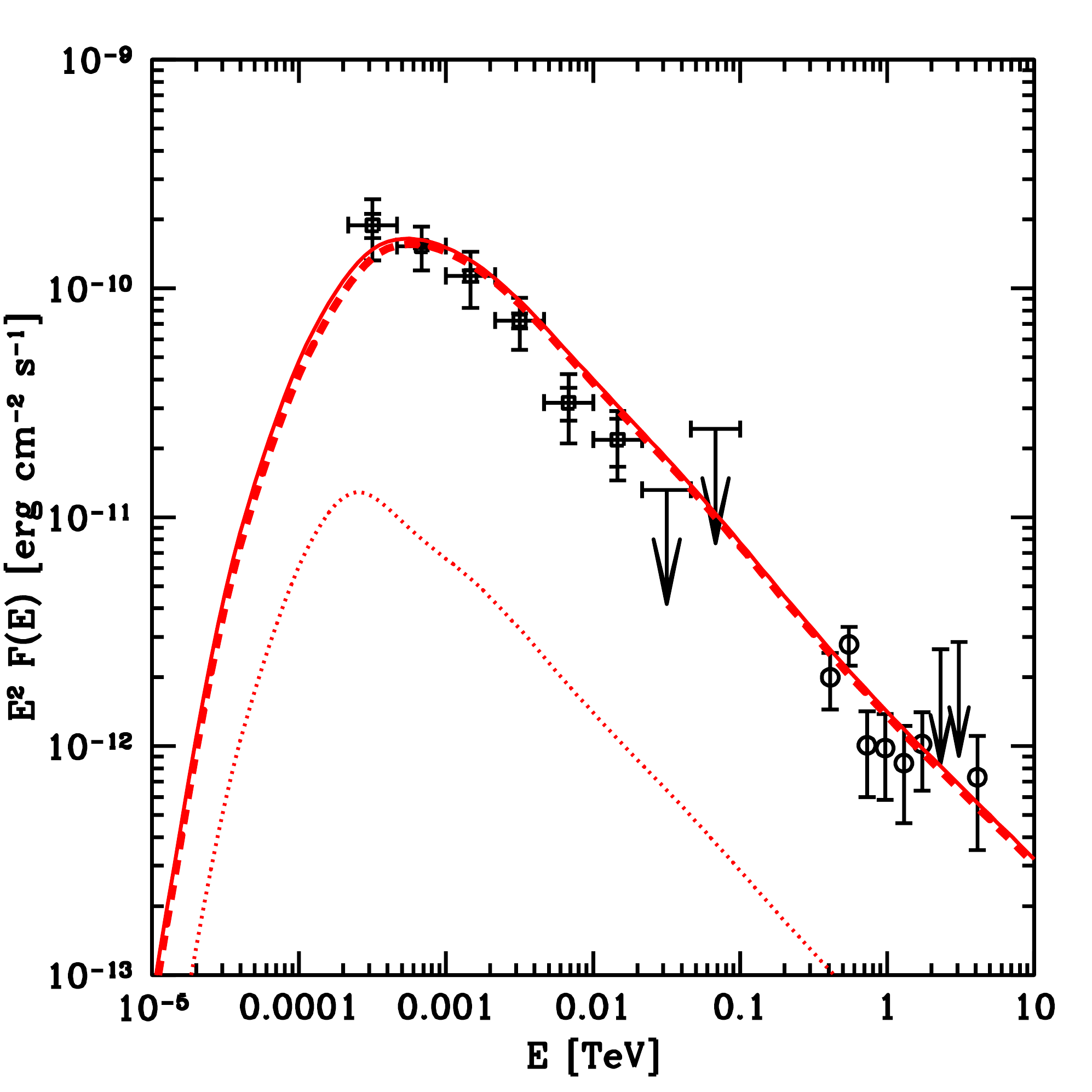}   
 \includegraphics[width=0.3\textwidth]{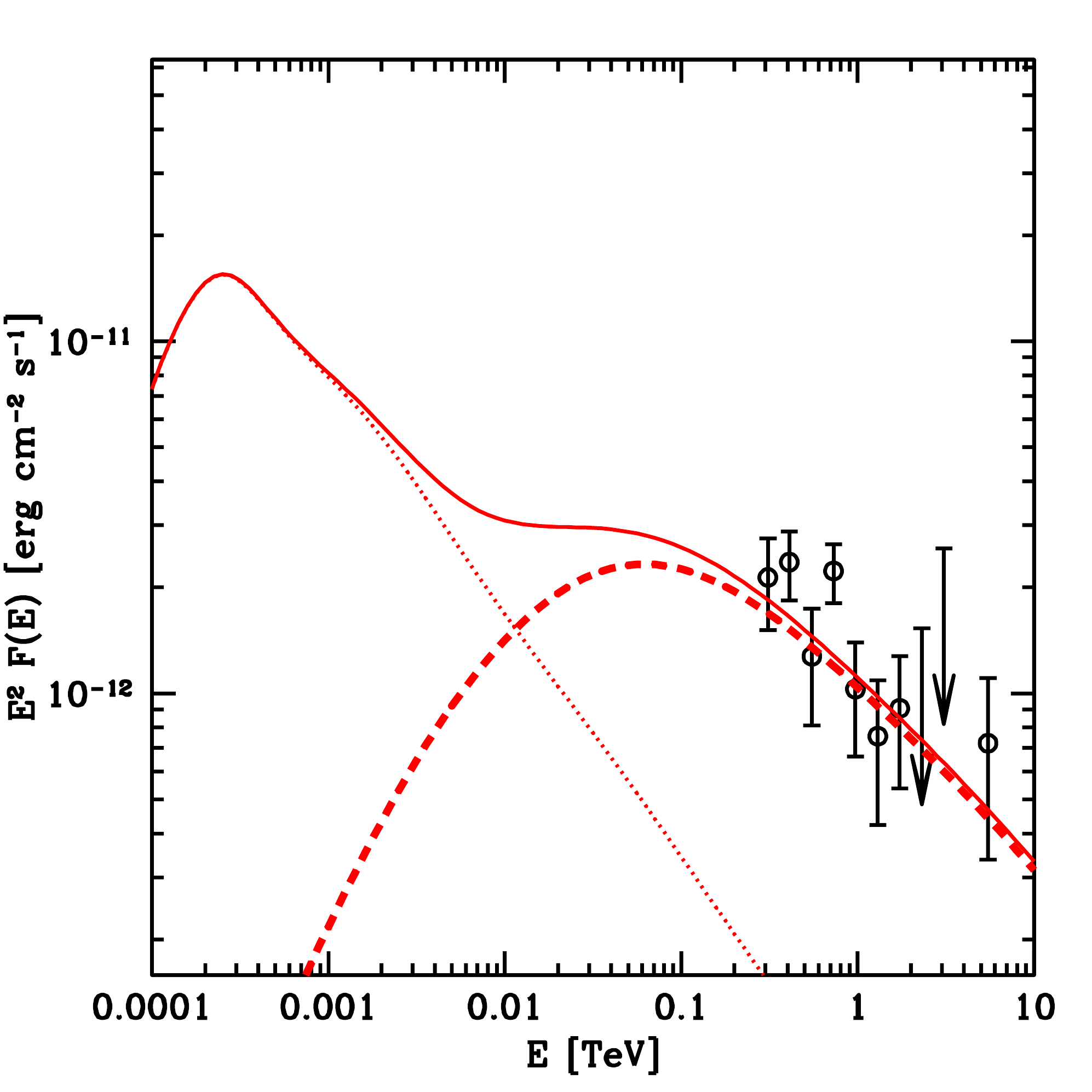}   
 \includegraphics[width=0.3\textwidth]{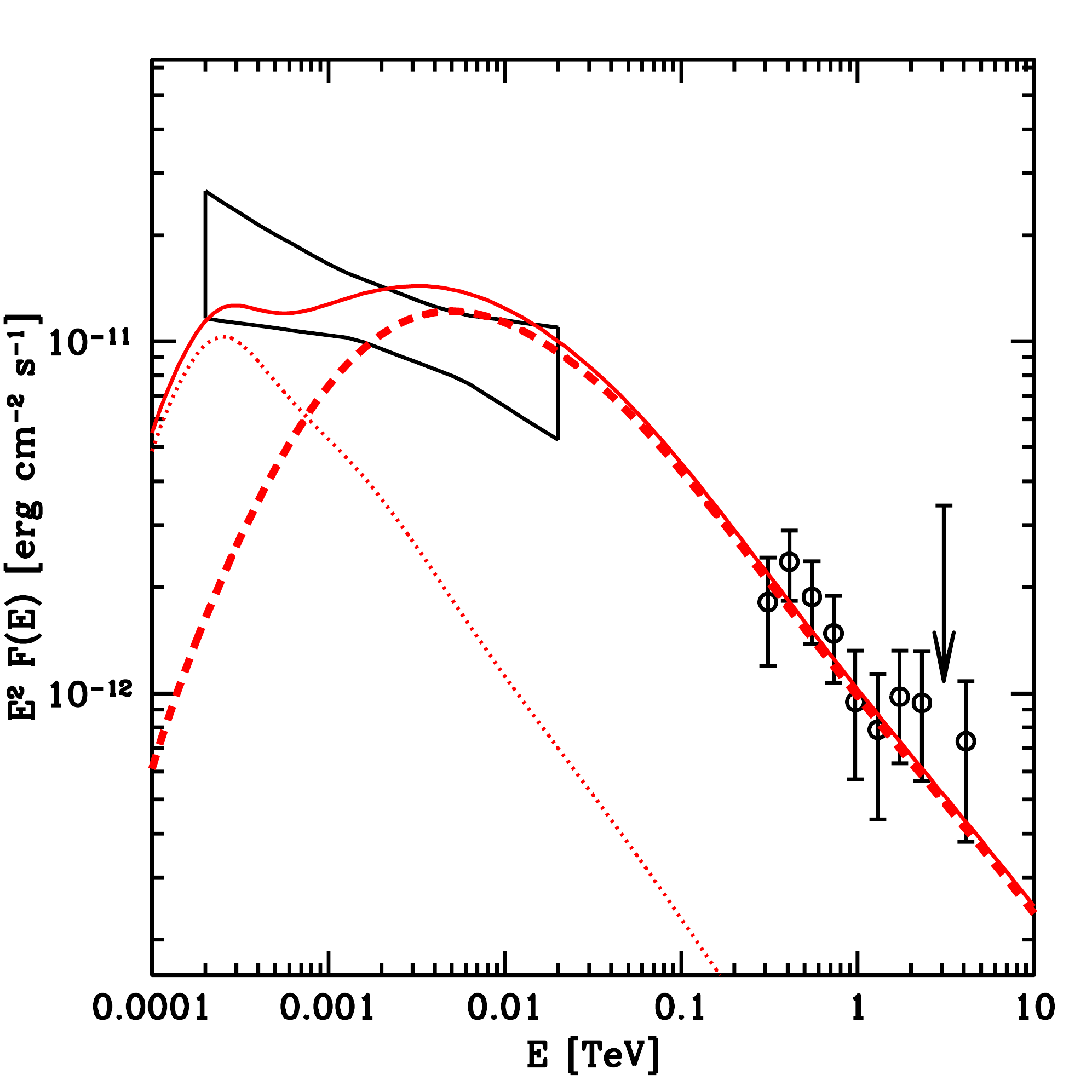}   
  \caption{Broad band fit to the gamma ray emission detected by FERMI and HESS from the sources HESS J1801-233, HESS J1800-240 A and B (left to right). Dashed lines represent the contribution to the gamma ray emission from CRs that escaped W28, dotted lines show the contribution from the CR galactic background, and solid lines the total emission. Distances to the SNR centre are 12, 65, and 32 pc (left to right). FERMI and HESS data points are plotted in black. No GeV emission has been detected from HESS J1800-240 A.}
  \label{fig2}
\end{figure}

Fig.~\ref{fig2} shows a fit to the broad band gamma ray spectrum measured from FERMI and HESS. The three panels refers to (left to right) the sources HESS J1801-233, HESS J1800-240 A and B, respectively. Dashed lines represent the contribution to the emission from CRs that escaped from W28, dotted lines the contribution from background CRs, and solid lines the total emission. Since FERMI data refers to the emission after background subtraction, dashed lines have to be compared with data points. The (often non-trivial) background subtraction issue might add another source of uncertainty in the comparison between data and predictions. An acceleration efficiency $\eta = 30\%$ and a diffusion coefficient $\chi = 0.06$ have been assumed, while the distance from the SNR centre is assumed to be (left to right) 12, 65, and 32 pc. Keeping in mind all the above mentioned caveats, it is encouraging to see that a qualitative agreement exists between data and predictions also in the GeV band.

We investigated the possibility that the gamma ray emission detected from the MCs in the region of the SNR W28 is produced by CRs that escaped the SNR. This interpretation requires the CR diffusion coefficient in that region to be significantly suppressed with respect to the average galactic one. Such suppression might be the result of an enhancement in the magnetic turbulence due to CR streaming away from the SNR.

\begin{acknowledgements}
SG thanks M. Walmsley for many interesting discussions. The research leading to these results has received funding from the European Union [FP7/2007-2013] under grant agreement n$^o$ 256464 and 221093.
\end{acknowledgements}

%
%
%
%
%

%

\begin{thebibliography}{}
\bibitem[Abdo et al.(2010)]{fermi} Abdo, A.A., et al. 2010, ApJ, 718, 348
\bibitem[Aharonian \& Atoyan(1996)]{atoyan} Aharonian, F.A., \& Atoyan, A. 1996, A\&A, 309, 917
\bibitem[Aharonian et al.(2008)]{hess} Aharonian, F.A., et al. 2008, A\&A, 481, 401
\bibitem[Berezinskii et al.(1990)]{CRbook} Berezinskii, V.S., Bulanov, V.A., Dogiel, V.L., Ginzburg, V.L., \& Ptuskin, V.S. 1990, Astrophysics of Cosmic Rays (Amsterdam: North-Holland)
\bibitem[Casanova et al.(2010)]{casanova2010} Casanova, S., et al. 2010, PASJ, in press -- arXiv:1003.0379
\bibitem[Cioffi et al.(1988)]{cioffi} Cioffi, D.F., McKee, C.F., \& Bertschinger, E. 1988, ApJ, 334, 252
\bibitem[Farmer \& Goldreich(2004)]{farmer} Farmer, A.J., \& Goldreich, P. 2004, ApJ, 604, 671
\bibitem[Fujita et al.(2009)]{fujita} Fujita, Y. Ohira, Y., Tanaka, S.J., \& Takahara, F. 2009, ApJ, 707, L179 
\bibitem[Gabici \& Aharonian(2007)]{gabici2007} Gabici, S., \& Aharonian, F.A. 2007, ApJ, 665, L131
\bibitem[Gabici et al.(2009)]{gabici2009} Gabici, S., Aharonian, F.A., \& Casanova, S. 2009, MNRAS, 396, 1629 (GAC2009)
\bibitem[Giuliani et al.(2010)]{agile} Giuliani, A., et al. 2010, A\&A, 516, L11
\bibitem[Kamae et al.(2006)]{kamae} Kamae, T., Karlsson, N., Mizuno, T., Abe, T., \& Koi, T. 2006, ApJ, 647, 692 
\bibitem[Kulsrud \& Pierce(1969)]{kulsrud} Kulsrud, R., \& Pearce, W.P. 1969, ApJ, 156, 445
\bibitem[Li \& Chen(2010)]{li} Li, H., \& Chen, Y. 2010, MNRAS, in press -- arXiv:1009.0894
\bibitem[Ohira et al.(2010)]{ohira} Ohira, Y., Murase, K., \& Yamazaki, R. 2010, MNRAS, in press -- arXiv:1007.4869
\bibitem[Ptuskin \& Zirakashvili(2005)]{vladvlad} Ptuskin, V.S., \& Zirakashvili, V.N. 2005, A\&A, 429, 755
\bibitem[Ptuskin et al.(2008)]{ptuskin} Ptuskin, V.S., Zirakashvili, V.N., \& Plesser, A.A. 2008, Adv. Space Res., 42, 486
\bibitem[Rho \& Borkowski(2002)]{rho} Rho, J., \& Borkowski, K.J. 2002, ApJ, 575, 201

\end{thebibliography}
\end{document}